\documentclass[%
reprint,
 showpacs,preprintnumbers,
 amsmath,amssymb,
 aps,
prb,
 floatfix,
]{revtex4-1}

\usepackage{booktabs}
\usepackage{array}
\newcolumntype{L}[1]{>{\raggedleft\arraybackslash}p{#1}}
\AtBeginDocument{
\heavyrulewidth=.08em
\lightrulewidth=.05em
\cmidrulewidth=.03em
\belowrulesep=.65ex
\belowbottomsep=0pt
\aboverulesep=.4ex
\abovetopsep=0pt
\cmidrulesep=\doublerulesep
\cmidrulekern=.5em
\defaultaddspace=.5em
}

\usepackage{color}
\usepackage{graphicx}
\usepackage{dcolumn}
\usepackage{bm}
\usepackage{cancel}
\usepackage[utf8]{inputenc}

\providecommand{\abs}[1]{\vert #1\vert}
\newcommand{\etal}{\textit{et al}.}
\newcommand{\ie}{\textit{i}.\textit{e}.}

\begin{document}

\title{Energetic analysis of disorder effects in an artificial spin ice with dipolar interactions}

\author{M. Di Pietro Mart\'inez}
\email{mdipietro@ifimar-conicet.gob.ar}
\author{R. C. Buceta}

\affiliation{Instituto de Investigaciones F\'isicas de Mar del Plata (CONICET -- UNMdP)}
\affiliation{Departamento de F\'isica, Facultad de Ciencias Exactas y Naturales,
             Universidad Nacional de Mar del Plata.\\
             De\'an Funes 3350, 7600 Mar del Plata, Argentina}

\date{\today}

\begin{abstract}
We study the effect of quenched disorder in square artificial spin ice by means of numerical simulations.
We introduce disorder in the length of magnetic islands using two kinds of distributions: Gaussian and uniform.
As the system behavior depends on its geometrical parameters, we focus on studying it in the proximity of the ice regime which is quite difficult to thermalize both in experiments and simulations.
We show how length disorder affect the antiferromagnetic and (locally) ferromagnetic ordering, by inducing the system, in the case of weak disorder, to intermediate or mix states.
Moreover, in the case of strong disorder, ferromagnetic plaquettes prevail regardless of whether the mean length of the islands corresponds to an antiferromagnetic ordering.
\end{abstract}

\maketitle

\section{Introduction}
\label{sec:intro}

Artificial spin ice (ASI) consists of a lithographically manufactured two-dimensional array of ferromagnetic nanoislands with a strong shape anisotropy resulting in single-domains that behave like giant Ising spins~\cite{Martin2003}.
In natural spin ices, such as, for example, the rare earth pyrochlore Ho$_2$Ti$_2$O$_7$, the local ordering of magnetic moments is particularly difficult to measure~\cite{Harris1997}.
But artificial frustrated magnetic systems make possible to access directly the degrees of freedom, \ie{} the spins~\cite{Ladak2010,Mengotti2011,morgan2011magnetic,zhang2012perpendicular,heyderman2013artificial,montaigne2014size,cumings2014focus,drisko2015fepd}.
This advantage allows to test and reproduce theoretical models, as well as to understand more about their three-dimensional analogs.
But even more importantly, it allows us to design a system rather than discover it~\cite{Wang2006}.

\begin{figure}
\centering
\resizebox{0.35\textwidth}{!}{
\begingroup%
  \makeatletter%
  \providecommand\color[2][]{%
    \errmessage{(Inkscape) Color is used for the text in Inkscape, but the package 'color.sty' is not loaded}%
    \renewcommand\color[2][]{}%
  }%
  \providecommand\transparent[1]{%
    \errmessage{(Inkscape) Transparency is used (non-zero) for the text in Inkscape, but the package 'transparent.sty' is not loaded}%
    \renewcommand\transparent[1]{}%
  }%
  \providecommand\rotatebox[2]{#2}%
  \newcommand*\fsize{\dimexpr\f@size pt\relax}%
  \newcommand*\lineheight[1]{\fontsize{\fsize}{#1\fsize}\selectfont}%
  \ifx\svgwidth\undefined%
    \setlength{\unitlength}{170.24042084bp}%
    \ifx\svgscale\undefined%
      \relax%
    \else%
      \setlength{\unitlength}{\unitlength * \real{\svgscale}}%
    \fi%
  \else%
    \setlength{\unitlength}{\svgwidth}%
  \fi%
  \global\let\svgwidth\undefined%
  \global\let\svgscale\undefined%
  \makeatother%
  \begin{picture}(1,1.10641993)%
    \lineheight{1}%
    \setlength\tabcolsep{0pt}%
    \put(0.41359619,0.46955652){\color[rgb]{0,0,0}\makebox(0,0)[lt]{\lineheight{0}\smash{\begin{tabular}[t]{l}Type III\end{tabular}}}}%
    \put(0,0){\includegraphics[width=\unitlength,page=1]{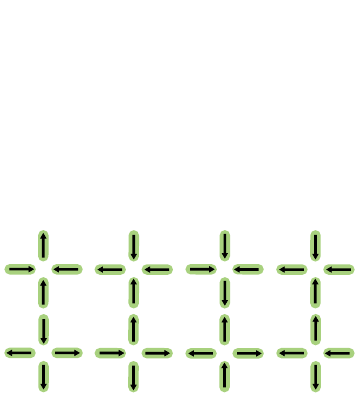}}%
    \put(0.04477754,1.00129834){\color[rgb]{0,0,0}\makebox(0,0)[lt]{\lineheight{0}\smash{\begin{tabular}[t]{l}Type I\end{tabular}}}}%
    \put(0.42175819,1.00129834){\color[rgb]{0,0,0}\makebox(0,0)[lt]{\lineheight{0}\smash{\begin{tabular}[t]{l}Type II\end{tabular}}}}%
    \put(0,0){\includegraphics[width=\unitlength,page=2]{fig1.pdf}}%
    \put(0.79207575,1.00402469){\color[rgb]{0,0,0}\makebox(0,0)[lt]{\lineheight{0}\smash{\begin{tabular}[t]{l}Type IV\end{tabular}}}}%
    \put(0,0){\includegraphics[width=\unitlength,page=3]{fig1.pdf}}%
    \put(0.2767923,1.06436689){\color[rgb]{0,0,0}\makebox(0,0)[lt]{\lineheight{0}\smash{\begin{tabular}[t]{l}Ice-like\end{tabular}}}}%
  \end{picture}%
\endgroup%
}
\caption{There are $16$ possible configurations for each vertex according to the orientation of its spins. Furthermore, these can be classified into four different topological groups. Type I and II satisfy the ice rule while Type III and IV do not. Type II and III vertices have a net magnetization.}
\label{fig:vertices}
\end{figure}

In 2006, Wang and collaborators created experimentally an ASI with ferromagnetic islands forming a square lattice~\cite{Wang2006}.
The magnetic force microscopy measurement of these samples showed a strong orientation of the magnetic moments in the longitudinal direction of each island; then, each island can be modeled as an in-plane spin.
The point where four islands concur is called a vertex.
For each vertex, there are $2^4$ possible configurations, according to the orientation of the four spins that form the vertex, which in turn can be grouped into four topologically different types (see Fig.~\ref{fig:vertices}).
Type I and II have two spins pointing inwards and two outwards of the vertex, meeting the so-called ice rule.
This rule was originally proposed by Pauling for the proton orderings in water ice, but a perfect mapping with these magnetic materials
was found later, hence the name \emph{spin ice}~\cite{Pauling1935,Harris1997,Petrenko1999,bramwell2001spin,Nisoli2013,Bramwell2013}.
The difference between these two types is that, unlike Type II vertices, Type I have null magnetization.
In contrast, Type III and IV vertices do not meet the ice rule; in this context, they are usually called \textit{defects}.
If the orientation of the spins were completely random, the following vertex population is expected according to the number of configuration each group has: $12.5$\% Type I, $25$\% Type II, $50$\% Type III and $12,5$\% Type IV.
At room temperature, Wang \etal{} measured that more than $70$\% of the vertices met the ice rule and that this percentage was reduced by increasing the lattice spacing~\cite{Wang2006}.
The greater the spacing, the more similar the situation to having a random configuration, which is equivalent to having non-interacting islands.
In this way, they managed to observe the spin-ice behavior in these artificial systems.

Inspired by these results, M\"oller and Moessner modeled this system and performed numerical simulations with an array of spins under dipolar interactions~\cite{Moller2006}.
They added a height separation between the spins oriented vertically and those oriented horizontally.
In a tetrahedrical lattice, such as the one found in natural spin ices, the distance between any pair of spins (belonging to the same tetrahedron) is the same, and therefore, the dipolar energy has the same value for any pair. On the contrary, in the square ASI, the distance between collinear spins is different from the distance between perpendicular spins. Hence, the need to add this height parameter.
In this way, this system is halfway between three-dimensional (natural) and two-dimensional (artificial) spin ices~\cite{Chern2014}.
Another lattice that is also studied in the cited article and in Ref.~\cite{Moller2009}, and which is usually used to model its three-dimensional counterpart, is the kagome. Unlike the square lattice, the distance between the spins of each plaquette in a kagome lattice is the same, so the height parameter is no longer necessary.

The study of these two geometries is very extensive, either through simulations or experiments, and allowed to understand more about spin ices and frustrated systems in bulk materials~\cite{Mol2009,Nisoli2010,Ladak2010,Arnalds2012,Nisoli2013,NisoliConference,canals2016fragmentation}.
In particular, it allowed to understand more about the interactions. Various models were studied where both short and long-ranged interactions were considered~\cite{Wills2002,Nisoli2007,Chern2011,Rougemaille2011}.

The effect of disorder is still an open question and a field to study in ASI~\cite{Nisoli2013,Bramwell2006,Moller2006,Bramwell2013}.
Budrikis \etal{} (2012) studied from the point of view of theory, simulations and experiments the influence of disorder on the response of the system to an external field~\cite{Budrikis2012}.
To do this, in the simulations, they proposed that each spin had an internal (coercive) field given by a Gaussian distribution. With this model, they managed to estimate the strength of disorder in a sample. There are several works that quantify the disorder~\cite{Ladak2010,Mengotti2011,Daunheimer2011,Pollard2012}, what is new in Budrikis' work is that they also study how disorder affects the dynamics of ASI.
Chern \etal{} (2014) take this same idea to model disorder and study the avalanches and critical behavior in square and kagome ASI when a magnetic field is applied in-plane~\cite{Chern2014b}.
This system shows a phase transition out-of-equilibrium induced by disorder strength.
Reichhardt \etal{} show that the avalanche distributions of this process
follow a power law~\cite{Reichhardt2015}.
Another way to include disorder in the system is to disconnect the islands by eliminating a given percentage of them at random. Greenberg \etal{} showed through simulations that this system changes its thermal behavior as the percentage of holes increases~\cite{Greenberg2018}. These results coincide with what was observed experimentally in diluted spin ice~\cite{Ke2007,Lin2014}, once again showing that the study of ASI allows modeling and better understanding of the behavior of spin ice.

The aim of this paper is to study, through simulations and an energetic analysis, the effect of geometrical disorder on an square ASI.
To do this, we propose to include disorder in the length of the islands.
In this way, the strength of disorder could be easily controlled and designed experimentally.

This article is organized as follows. Firstly, in Sec.~\ref{sec:model}, we present the details about the model and the way to introduce disorder.
We calculate the energy for this system and analyze the relationship between the strength of disorder and the spin interactions.
In Sec.~\ref{sec:results}, we show the results of numerical simulations, first, without disorder, to verify with previous results, and then, with disorder. We present the effect on thermodynamic regimes and analyze the results by studying how the energy contained in each type of vertex changes with the strength of disorder.
Also, we characterize the spin dynamics under the effect of disorder.
Finally, in Sec.~\ref{sec:conclusions}, we present the conclusions.

\section{Modeling an artificial spin-ice}
\label{sec:model}

\subsection{Description of the system}

\begin{figure}
\centering
\resizebox{0.3\textwidth}{!}{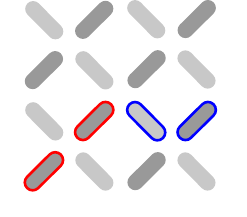}
\caption{Square ASI array with dipolar interactions: $a$ is the lattice spacing and $d$ is the island length; the width is considered negligible. There is a height gap $h$ between the position of islands which are oriented in perpendicular directions; this means that dark-gray shaded islands are located at height $z=h$ while light-gray shaded islands are located at height $z=0$. Each spin ${\bf S}_{\kappa}=\mu {\bf \hat{S}_{\kappa}}$ is considered as a magnetic dipole, with uniform magnetic density, whose magnetic field is equivalent to that of two effective charges $\pm q_{\kappa}=\mu/d_{\kappa}$ located one at each end of the island. Two islands which are first nearest-neighbors are marked in blue, and two islands which are second nearest-neighbors are marked in red.}
\label{fig:lattice}
\end{figure}

The square ASI is formed by islands of length $d$ arranged as shown in Fig.~\ref{fig:lattice}; the lattice spacing is $a$.
In this model, the width of the islands is considered negligible.
In turn, as said in the previous section,
we consider a height gap $h$ between the spins oriented in different directions.
In the figure, this is marked with different shades of gray.
In the experiments, the islands are formed by a ferromagnetic material and due to the strong shape anisotropy, the magnetization of these is forced to align along the easy axis. In this way, the islands behave effectively like Ising spins.
Then, each magnetic moment can have only two possible orientations.

\subsection{Hamiltonian}
\label{sec:calcenergia}

The system is described by the Hamiltonian
\begin{equation}
 H = \frac{1}{2} \sum_{\alpha, \beta (\alpha \neq \beta)} \varepsilon_{\alpha \beta},
 \label{eq:hamiltoniano}
\end{equation}
where $\varepsilon_{\alpha \beta}$ is the interaction energy between two spins and the sum is done over all pairs in the lattice.
To calculate $\varepsilon_{\alpha \beta}$, we take the spin ${\bf S}_{\kappa}=\mu {\bf \hat{S}_{\kappa}}$ as a magnetic dipole with uniform magnetic density.
This needle-shaped islands create a field equivalent to that of two effective charges $\pm q_{\kappa}=\mu/d_{\kappa}$ located one at each end of the island, ${\bf r}_{\kappa}^+$ and ${\bf r}_{\kappa}^-$, and separated a distance $d_{\kappa}$~\cite{Moller2009} (see Fig.~\ref{fig:lattice}).
Then, the potential created by a magnetic dipole is simply
\begin{equation}
\Phi_{\kappa}({\bf r}) = \frac{q_{\kappa}}{4\pi \epsilon_{0}} \left( \frac{1}{\abs{{\bf r}-{\bf r}_{\kappa}^+}}-\frac{1}{\abs{{\bf r}-{\bf r}_{\kappa}^-}}  \right).
\label{eq:dip0}
\end{equation}
Then, the interaction energy between two spins ${\bf S}_\alpha$ and ${\bf S}_\beta$ is
\begin{equation}
\varepsilon_{\alpha \beta} = q_{\beta} \bigl[ \Phi_\alpha({\bf r}_\beta^+)-\Phi_\alpha({\bf r}_\beta^-) \bigr].
\label{eq:dip1}
\end{equation}
Replacing the equation \eqref{eq:dip0} in \eqref{eq:dip1} and taking into account that ${\bf r}_{\kappa}^\pm = {\bf r}_{\kappa} \pm \frac{1}{2}\,d_{\kappa}\,{\bf \hat{S}_{\kappa}}$ (with $\kappa=\alpha,\beta$) and ${\bf r}_{\alpha\beta}={\bf r}_\beta-{\bf r}_\alpha$, the equation can be rewritten as
\begin{equation}
\begin{split}
 \varepsilon_{\alpha \beta} = \frac{\mathcal{D}}{d_{\alpha} d_{\beta}} & \left(  \frac{1}{\abs{{\bf r}_{\alpha\beta} + \frac{1}{2}(d_\beta{\bf \hat{S}_{\beta}}-d_\alpha{\bf \hat{S}_{\alpha}})}} \right.\\
 &+ \frac{1}{\abs{{\bf r}_{\alpha\beta} -\frac{1}{2}(d_\beta{\bf \hat{S}_{\beta}}- d_\alpha{\bf \hat{S}_{\alpha}})}} \\
 &- \frac{1}{\abs{{\bf r}_{\alpha\beta} + \frac{1}{2}(d_\beta{\bf \hat{S}_{\beta}} + d_\alpha{\bf \hat{S}_{\alpha}})}} \\
 &- \left. \frac{1}{\abs{{\bf r}_{\alpha\beta} - \frac{1}{2}(d_\beta{\bf \hat{S}_{\beta}} + d_\alpha{\bf \hat{S}_{\alpha}})}} \right),
\end{split}
\label{eq:dip2}
\end{equation}
where $\mathcal{D}=\mu_{0}\mu^{2}/4\pi$.
It can be shown that the limit $d_\kappa\rightarrow 0$ ($\kappa=\alpha,\beta$) for this expression is
\begin{equation}
 \lim_{d\rightarrow0} \varepsilon_{\alpha \beta} = \mathcal{D} \left( \frac{{\bf \hat{S}_{\alpha}} \cdot {\bf \hat{S}_{\beta}}}{r^{3}_{\alpha \beta}} - 3 \frac{({\bf \hat{S}_{\alpha}} \cdot {\bf \hat{r}_{\alpha \beta}})({\bf \hat{S}_{\beta}} \cdot {\bf \hat{r}_{\alpha \beta}})}{r^{5}_{\alpha \beta}}  \right),
 \label{eq:limite}
\end{equation}
which corresponds to the energy for point-like magnetic dipoles.

To study the effect of disorder, we consider that the length of each island $d_{\kappa}$
is given by a Gaussian distribution with mean $d$ and standard deviation $\sigma$, which determines the disorder strength.
The distribution is cut to $d_\kappa<a$ to avoid island overlapping.
To implement strong disorder, we use an uniform distribution with interval $[-\Delta,\Delta]$.

\subsection{First and second nearest-neighbors bond energy}

While the interactions considered in this model are long-ranged (see Eq.~\eqref{eq:dip2}), a good prediction of the ground state can be made by analyzing only the bond energies of first and second nearest-neighbors, $J_1$ and $J_2$, respectively~\cite{Moller2006,perrin2016extensive}.
In Fig.~\ref{fig:lattice}, the two islands marked in red are first nearest-neighbors and the two blue ones are second nearest-neighbors.
Replacing ${\bf \hat{S}_{\alpha}}=(0,1,0)$, ${\bf \hat{S}_{\beta}}=(1,0,0)$ and ${\bf r_{\alpha \beta}}=(a/2,a/2,h)$ in Eq.~\eqref{eq:dip2}, the bond energy of two first nearest-neighbor spins is
\begin{equation}
\begin{split}
 J_{1} & = \frac{\mathcal{D}}{d^2} \left[~ |( a/2+d/2, a/2-d/2, h)|^{-1}  \right. \\
 & + |( a/2-d/2, a/2+d/2, h)|^{-1} \\
 & - |( a/2+d/2, a/2+d/2, h)|^{-1} \\
 & \left. - |( a/2-d/2, a/2-d/2, h)|^{-1} \right]\\
 & = \frac{\mathcal{D}}{a\,d^2} \left[ 2 \left( \frac{1 + (d/a)^{2}}{2} + (h/a)^{2} \right)^{-1/2} \right. \\
 & - \left( \frac{(1 + d/a)^{2}}{2} + (h/a)^{2} \right)^{-1/2} \\
 & - \left. \left( \frac{(1 - d/a)^{2}}{2} + (h/a)^{2} \right)^{-1/2} \right].
\end{split}
\end{equation}
While, replacing ${\bf \hat{S}_{\alpha}}=(0,1,0)$, ${\bf \hat{S}_{\beta}}=(0,1,0)$ and ${\bf r_{\alpha \beta}}=(0,a,0)$ in Eq.~\ref{eq:dip2}, the bond energy of two second nearest-neighbors is
\begin{equation}
\begin{split}
 J_{2} &= \frac{\mathcal{D}}{d^2} \left( \frac{2}{a}-|( 0, a+d, 0)|^{-1}- |( 0, a-d, 0)|^{-1} \right)\\
 &= \frac{\mathcal{D}}{a\,d^2} \left( 2 -|1+d/a|^{-1}- |1-d/a|^{-1} \right).
\end{split}
\end{equation}

\begin{figure*}
\centering
\includegraphics[width=0.7\linewidth]{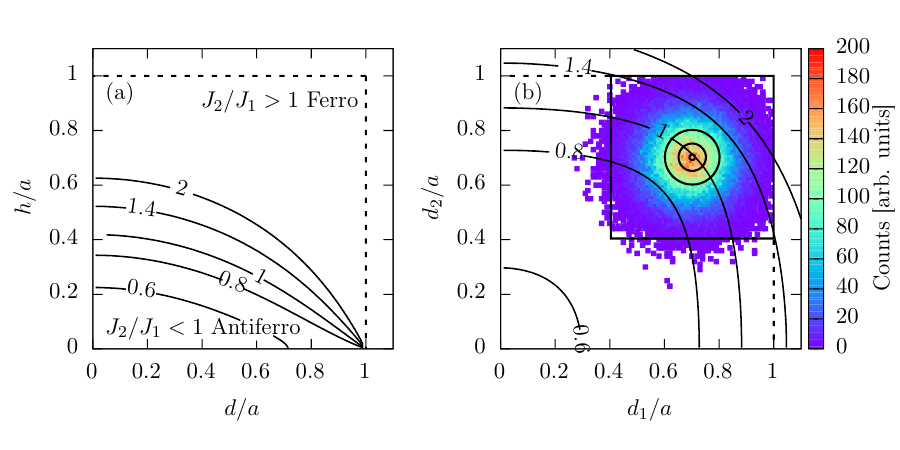}
\caption{(a) Contour plots for the ratio $J_{2}/J_{1}$ (Eq.~\eqref{eq:j2j1Moller}) as a function of the geometrical parameters when there is no disorder in the lattice; these results match the ones from Ref.~\cite{Moller2006}. When $J_{2}/J_{1}<1$, Type I vertices have a lower energy and so the antiferromagnic ordering is expected. While, when $J_{2}/J_{1}>1$, Types II vertices have lower energy and so the (locally) ferromagnetic ordering is predicted.
(b) Contour plots for the ratio $J_{2}/J_{1}$ when island lengths $d_1$ and $d_2$ are different (Eqs.~\eqref{eq:j1} and \eqref{eq:j2}).
The length of each island is given by a Gaussian distribution with mean $d$ and standard deviation $\sigma$, or by a uniform distribution. The color map (a histogram of lengths) was obtained for $d=0.702$ and $\sigma=0.1$. The circles indicate where the first $\sigma$ of the distribution would be if the deviation were $0.01$, $0.05$ or $0.1$. The case of uniform disorder, with $\Delta=0.298$, is indicated with a solid-line rectangle. In addition, the restriction $d_k<a$ imposed to avoid island overlapping is indicated with dashed lines.}
\label{fig:j2j1}
\end{figure*}

If the ratio $J_{2}/J_{1} < 1$, Type I vertices are expected to be favored since they have lower energy ($-J {\bf \hat{S}_{\alpha}} \cdot {\bf \hat{S}_{\beta}}$).
While if $J_{2}/J_{1}>1$, Type II vertices will prevail.
In Fig.~\ref{fig:j2j1}~(a), contour plots are shown for the expression
\begin{widetext}
\begin{equation}
 J_{2}/J_{1} = \frac{2 -|1+d/a|^{-1}- |1-d/a|^{-1}}{ 2 \left( \frac{1 + (d/a)^{2}}{2} + \left(\frac{h}{a}\right)^{2} \right)^{-1/2} - \left( \frac{(1 + d/a)^{2}}{2} + \left(\frac{h}{a}\right)^{2} \right)^{-1/2} - \left( \frac{(1 - d/a)^{2}}{2} + \left(\frac{h}{a}\right)^{2} \right)^{-1/2}},
\label{eq:j2j1Moller}
\end{equation}
\end{widetext}
as a function of the geometrical parameters $h$ and $d$. By appropriately selecting these parameters, it can be obtained an antiferromagnetic system, where all the vertices are Type I, or a (locally) ferromagnetic system, where all the vertices are Type II.
In their article, M\"oller \etal{} obtains an antiferromagnetic system for the parameters $h=0.205$ and $d=0.7$, and a ferromagnetic for $h=0.207$ and $d=0.7$ (with $a=1$).
The goal in choosing these values, where $J_2/J_1$ is close to $1$, is to evidence the ferromagnetic-antiferromagnetic transition.
Nevertheless, tuning the parameters so that Type I and II vertices have exactly the same energy (ice-like regime) is very difficult.

Eq.~\eqref{eq:j2j1Moller} was obtained considering a system where all islands have the same length $d$.
If the lengths are different, the equations for $J_1$ and $J_2$ take the form
\begin{equation}
\begin{split}
J_1 &= \frac{\mathcal{D}}{a d_1 d_2}\left( \left(\frac{(1+d_2/a)^2}{4}+\frac{(1-d_1/a)^2}{4}+(h/a)^2\right)^{-1/2} \right.\\
&+\left(\frac{(1-d_2/a)^2}{4}+\frac{(1+d_1/a)^2}{4}+(h/a)^2\right)^{-1/2} \\
&- \left(\frac{(1+d_2/a)^2}{4}+\frac{(1+d_1/a)^2}{4}+(h/a)^2\right)^{-1/2}\\
&-\left. \left(\frac{(1-d_2/a)^2}{4}+\frac{(1-d_1/a)^2}{4}+(h/a)^2\right)^{-1/2} \right),
\end{split}
\label{eq:j1}
\end{equation}
\begin{equation}
\begin{split}
J_2 = \frac{\mathcal{D}}{a d_1 d_2} & \left( \left|1 + \frac{d_2-d_1}{2a}\right|^{-1} + \left|1 - \frac{d_2-d_1}{2a}\right|^{-1} \right.\\
&- \left. \left|1+\frac{d_2+d_1}{2a}\right|^{-1} - \left|1-\frac{d_2+d_1}{2a}\right|^{-1} \right).
\end{split}
\label{eq:j2}
\end{equation}

In Fig.~\ref{fig:j2j1}~(b), contour plots for the ratio $J_2/J_1$ according to Eqs.~\eqref{eq:j1} and \eqref{eq:j2} are shown, as a function of the lengths $d_1$ and $d_2$;
the height gap is $h=0.205$.
In the following, without loss of generality, we fixed $a=1$.
The distribution of island lengths for Gaussian disorder with mean $d=0.702$ and standard deviation $\sigma=0.1$ is shown as a color map.
Note that to avoid overlapping, the length is restricted to $d_k<a$; this is marked with a dashed line.
Circles indicate the first $\sigma$ region of the distribution for $\sigma=0.01$, $0.05$ and $0.1$,
while the rectangle indicates the amplitude of the uniform distribution with $\Delta=0.298$.
These are the disorder strengths that we study in this paper.

\section{Disorder effects}
\label{sec:results}

\subsection{Thermodynamic regimes}

\begin{figure*}
\centering
\includegraphics[width=\linewidth]{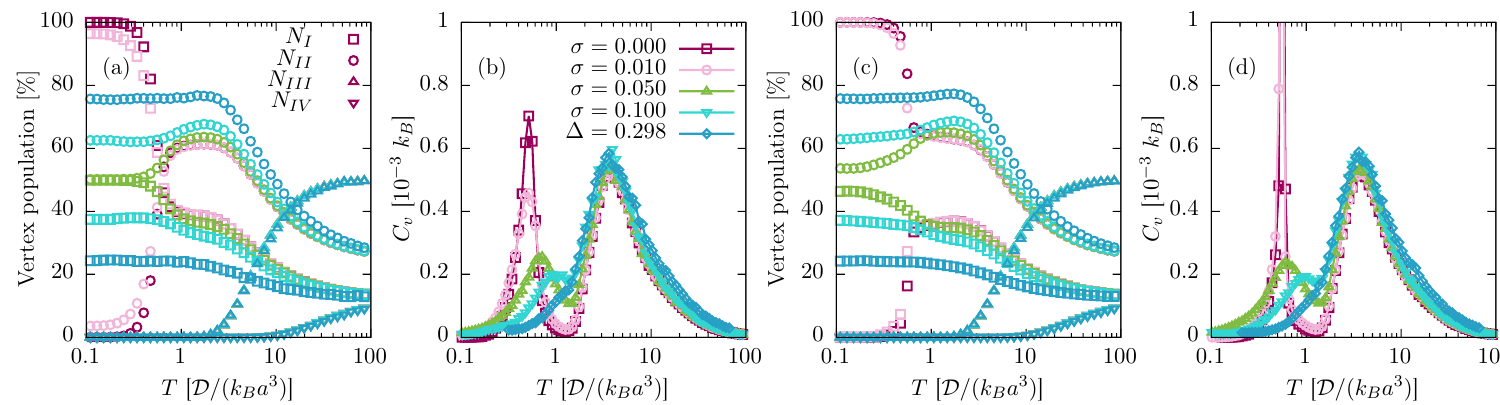}
\caption{Plots (a) and (c) show the population for each type of vertex, as a function of the temperature $T$, and plots (b) and (d) show the specific heats as functions of the temperature $T$, for different disorder strengths $\sigma$, or $\Delta$ in case of uniform disorder. Simulation parameters: $N=576$, $r_{cut}=4$, $a=1$, $h=0.205$ and $d=0.702$ for (a) and (b), or $d=0.704$ for (c) and (d).}
\label{fig:verticespop}
\end{figure*}

In this section, we present the results of numerical simulations with different disorder strengths; see Appendix~\ref{app:compdetails} for computational details.
In Fig.~\ref{fig:verticespop}, we show the vertex population and the specific heat as functions of the temperature $T$ for a system with $h=0.205$ and mean length $d=0.702$ (plots (a) and (b)) or $d=0.704$ (plots (c) and (d)).

Let us analyze first the system without disorder, $\sigma=0$, whose behavior is known, starting from the high temperature regime.
This corresponds, both for $d=0.702$ and $d=0.704$, to the random state
where the vertex population is given
according to the number of configurations each group has (see Fig.~\ref{fig:vertices}).
The specific heat $C_v$ shows, for both systems, an increase for $T\approx4$ due to the disappearance of Type III vertices.
There is a crossover between the random state and the ice regime,
where there are only ice-type vertices, \ie{} Type I and II.
This crossover is also characterized by a the decrease to zero of the single-spin-flip algorithm efficiency, which shows the need to use other algorithm, as discused in Appendix~\ref{app:compdetails}.
Then, as the temperature decreases, a phase transition at $T\approx0.5$ appears, evidenced by the specific heat peak observed both in plots (b) and (d).
Finally, the degeneracy is lift and Type I vertices prevail for $d=0.702$, while Type II predominates for $d=0.704$.
Then, it is said that for $d=0.702$ the ground state is antiferromagnetic, while for $d=0.704$ it is ferromagnetic.
These results agree with those already known and published~\cite{Moller2006,Moller2006t,Thonig2014}. Let us now analyze the effect of disorder.

First, it can be observed that disorder, regardless of its intensity, does not affect the random behavior at high temperatures, as well as the behavior of Type III and IV vertices at all temperatures.
However, it does affect Type I and II for lower temperatures.
In the antiferromagnetic case, we found that weak disorder, $\sigma=0.01$, allows a small percentage of Type II vertices at the ground state.
As disorder increases, the amount of Type I vertices reduces and so Type II population grows, and, at the same time, the phase transition peak dissolves.
A priori, one could expect this to continue until reaching the ice regime. However, for maximum disorder a different ground state is found where there are $25$\% Type I vertices and $75\%$ Type II.
And, what is most striking is that for the ferromagnetic case the same sequence occurs:
As disorder increases, the population of Type II vertices reduces at first until it reaches the $50$--$50$\% ground state. And then, it grows again until the state where $25$\% of the vertices are Type I and $75\%$ are Type II.

Simulation results show that, with disorder, the completely antiferromagnetic or ferromagnetic ground states are lost, and intermediate regimes are found instead.
This is also observed in the specific heat, in which the phase transition peak decreases in intensity until is lost, while, at the same time, it shifts to higher temperatures.
A relevant result is that disorder strength can be tuned to obtain the ice regime, for which the energies of the Type I and II vertices are equal.
For $\sigma\leq0.05$, population changes until it reaches the $50$--$50$ state.
In the case of maximum possible disorder, that is, for uniform disorder with $\Delta=0.298$ (or $\Delta=0.296$ for $d=0.704$),
the final state is the same for both systems.
Regardless of whether the system was originally (\ie{} without disorder) antiferromagnetic or ferromagnetic, for strong disorder a low temperature regime is obtained where approximately $3/4$ of the vertices are Type II and $1/4$ are Type I.

\begin{figure}
\centering
\includegraphics[width=\linewidth]{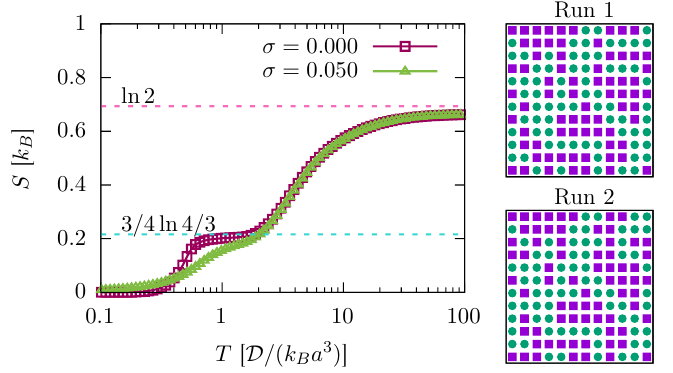}
\caption{Entropy as a function of the temperature, with and without disorder. Two different runs for the same disorder realization are shown ($\sigma=0.05$). Simulation parameters: $N=576$, $r_{cut}=4$, $a=1$, $h=0.205$ and $d=0.702$.}
\label{fig:entropy}
\end{figure}

Another question that rises is whether the thermodynamic regimes have an additional frustration created by disorder.
To study this, we perform different runs with the same seed for generating the island lengths, \ie{} the same realization of disorder, but with different initial spin configurations. After relaxation, we found that the samples evolve not only to the same percentage of Type I and II vertices as expected, but also to the same vertex configuration. This means that a particular vertex always finds the same state related to the realization of disorder. In Fig.~\ref{fig:entropy}, we show an example of two runs for $h=0.205$, $d=0.702$ and $\sigma=0.05$ with a $95$\% of overlapping.

Furthermore, if we measure the entropy, we found that there is no residual entropy (see Fig.~\ref{fig:entropy}).
When there is no disorder in the sample, the high-temperature entropy equals $\ln 2$, which is related to the $2^n$ possible configurations. Then, there is a plateau related to the ice regime, where each vertex can be Type I or II indistinctly. Finally, for lower temperatures, the frustration is lifted as the system orders antiferromagnetically or ferromagnetically according to the chosen island length $d$, hence there is no residual entropy.
When introducing disorder, we observe that the step related to the phase transition becomes less steep, which agrees with the decreasing of the pick in the heat capacity, but there is no residual term related to disorder.

\subsection{Vertex energy}
\label{sec:vertexenergy}

To analyze the regimes from an energetic point of view, we calculate the energy contained in each type of vertex and the dispersion in their values caused by disorder.
The occurrence of different types of vertices will be linked to these results.
To do this, we consider a vertex formed by islands of lengths $d_1$, $d_2$, $d_3$ and $d_4$, located at ${\bf r}_{1}=(0,0,0)$, ${\bf r}_{2}=(a/2,a/2,h)$, ${\bf r}_{3}=(0,a,0)$ and ${\bf r}_{4} = (-a/2,a/2,h)$, respectively.
The interaction energy between any pair of islands of the vertex, ${\bf S_{\alpha}}$ and ${\bf S_{\beta}}$, separated a distance ${\bf r}_{\alpha \beta}=\mathbf{r}_\beta-\mathbf{r}_\alpha$ is given by Eq.~\ref{eq:dip2}, where $\varepsilon_{\alpha \beta}=\varepsilon({\bf r}_{\alpha \beta},{\bf \hat{S}_{\alpha}},{\bf \hat{S}_{\beta}},d_\alpha,d_\beta$).
Then, the energy contained in a vertex is
$E\bigl[\hat{\mathbf{S}}_1,\hat{\mathbf{S}}_2,\hat{\mathbf{S}}_3,\hat{\mathbf{S}}_4\bigr]=\frac{1}{2}\sum_{\substack{\alpha,\beta=1\\ \alpha\neq\beta}}^4 \varepsilon_{\alpha\beta}$.
Actually, $E$ is also a function of the lengths $d_1$, $d_2$, $d_3$ and $d_4$, but we omit it to lighten the notation.
We can obtain the energy for each type of vertex $E_{\mbox{Type}}$ by replacing $\hat{S}$ by the corresponding values.
Then, the energy of a Type I vertex is calculated as
\begin{equation}
\begin{split}
E_\mathrm{I} & = \frac{1}{2} \bigl\{E\bigl[(0, -1, 0),(-1, 0, 0),(0, 1, 0),(1, 0, 0)\bigr] \bigr.\\
    & + \bigl. E\bigl[(0, 1, 0),(1, 0, 0),(0, -1, 0),(-1, 0, 0)\bigr]\bigr\},
\end{split}
\end{equation}
where we averaged over the two configurations that belong to the same topological group (see Fig.~\ref{fig:vertices}).
Solving the previous equation, we get
\begin{widetext}
\begin{equation}
\begin{split}
  E_\mathrm{I} =\sum_{i,j=\{(1,2),(1,4),(2,3),(3,4)\}} & \left[ - \left(d_i d_j \sqrt{h^2+\frac{d_j^2}{4}+\frac{a d_j}{2}+\frac{d_i^2}{4}+\frac{a d_i}{2}+\frac{a^2}{2}}\right)^{-1} \right.\\
  &+ \left(d_i d_j \sqrt{h^2+\frac{d_j^2}{4}+\frac{a d_j}{2}+\frac{d_i^2}{4}-\frac{a d_i}{2}+\frac{a^2}{2}}\right)^{-1}\\
  &+ \left(d_i d_j \sqrt{h^2+\frac{d_j^2}{4}-\frac{a d_j}{2}+\frac{d_i^2}{4}+\frac{a d_i}{2}+\frac{a^2}{2}}\right)^{-1}\\
  &- \left. \left(d_i d_j \sqrt{h^2+\frac{d_j^2}{4}-\frac{a d_j}{2}+\frac{d_i^2}{4}-\frac{a d_i}{2}+\frac{a^2}{2}}\right)^{-1}\right]\\
  + \sum_{i,j=\{(2,4),(1,3)\}} & \left[ \left( d_i d_j \left| \frac{d_i}{2} + \frac{d_j}{2} + a \right| \right)^{-1}
  +\left( d_i d_j \left| \frac{d_i}{2} + \frac{d_j}{2} - a \right| \right)^{-1} \right.\\
  &- \left. \left( d_i d_j \left| \frac{d_j}{2} - \frac{d_i}{2} + a \right| \right)^{-1}
  -\left( d_i d_j \left| \frac{d_j}{2} - \frac{d_i}{2} - a \right| \right)^{-1} \right].
\end{split}
\label{eq:energia1}
\end{equation}
\end{widetext}

Similarly, for the other types we have that
\begin{widetext}
\begin{equation}
\begin{split}
 E_\mathrm{II} &= \frac{1}{4} \left\lbrace E\left[(0,1,0),(1,0,0),(0,1,0),(1,0,0)\right] \right.
 + E\left[(0,1,0),(-1,0,0),(0,1,0),(-1,0,0)\right] \\
 &+ E\left[(0,-1,0),(1,0,0),(0,-1,0),(1,0,0)\right] 
 + \left. E\left[(0,-1,0),(-1,0,0),(0,-1,0),(-1,0,0)\right] \right\}\\
  &= \sum_{i,j=\{(2,4),(1,3)\}} \left[ - \left( d_i d_j \left| \frac{d_j}{2} + \frac{d_i}{2} + a \right| \right)^{-1} \right.
   - \left( d_i d_j \left| \frac{d_j}{2} + \frac{d_i}{2} - a \right| \right)^{-1} \\
   &+ \left( d_i d_j \left| \frac{d_j}{2} - \frac{d_i}{2} + a \right| \right)^{-1}
    + \left. \left( d_i d_j \left| \frac{d_j}{2} - \frac{d_i}{2} - a \right| \right)^{-1} \right],
 \end{split}
\label{eq:energia2}
\end{equation}

\begin{equation}
\begin{split}
 E_\mathrm{III} &= \frac{1}{8} \left\{ E\left[(0,1,0),(-1,0,0),(0,1,0),(1,0,0)\right] \right.
 +E\left[(0,-1,0),(1,0,0),(0,1,0),(1,0,0)\right] \\
 &+ E\left[(0,1,0),(1,0,0),(0,-1,0),(1,0,0)\right] 
 + E\left[(0,1,0),(1,0,0),(0,1,0),(-1,0,0)\right] \\
 &+ E\left[(0,-1,0),(-1,0,0),(0,-1,0),(1,0,0)\right] 
 +E\left[(0,-1,0),(-1,0,0),(0,1,0),(-1,0,0)\right] \\
 &+ E\left[(0,1,0),(-1,0,0),(0,-1,0),(-1,0,0)\right] 
 + \left. E\left[(0,-1,0),(1,0,0),(0,-1,0),(-1,0,0)\right] \right\}\\
 &= 0,
\end{split}
\label{eq:energia3}
\end{equation}

\begin{equation}
 \begin{split}
 E_\mathrm{IV} &= \frac{1}{2} \left\lbrace E\left[(0,-1,0),(1,0,0),(0,1,0),(-1,0,0)\right] + E\left[(0,1,0),(-1,0,0),(0,-1,0),(1,0,0)\right] \right\}\\
  & =\sum_{i,j=\left\lbrace(1,2),(1,4),(2,3),(3,4)\right\rbrace} \left[ \left(d_i d_j \sqrt{h^2+\frac{d_j^2}{4}+\frac{a d_j}{2}+\frac{d_i^2}{4}+\frac{a d_i}{2}+\frac{a^2}{2}}\right)^{-1} \right.\\
  &- \left(d_i d_j \sqrt{h^2+\frac{d_j^2}{4}+\frac{a d_j}{2}+\frac{d_i^2}{4}-\frac{a d_i}{2}+\frac{a^2}{2}}\right)^{-1}
  - \left(d_i d_j \sqrt{h^2+\frac{d_j^2}{4}-\frac{a d_j}{2}+\frac{d_i^2}{4}+\frac{a d_i}{2}+\frac{a^2}{2}}\right)^{-1}\\
  &+ \left. \left(d_i d_j \sqrt{h^2+\frac{d_j^2}{4}-\frac{a d_j}{2}+\frac{d_i^2}{4}-\frac{a d_i}{2}+\frac{a^2}{2}}\right)^{-1}\right]\\
  &+ \sum_{i,j=\{(2,4),(1,3)\}} \left[ \left( d_i d_j \left| \frac{d_i}{2} + \frac{d_j}{2} + a \right| \right)^{-1}
  +\left( d_i d_j \left| \frac{d_i}{2} + \frac{d_j}{2} - a \right| \right)^{-1} \right.\\
  &- \left. \left( d_i d_j \left| \frac{d_j}{2} - \frac{d_i}{2} + a \right| \right)^{-1}
  -\left( d_i d_j \left| \frac{d_j}{2} - \frac{d_i}{2} - a \right| \right)^{-1} \right].
 \end{split}
\label{eq:energia4}
\end{equation}
\end{widetext}

\begin{figure}[!htb]
\centering
\includegraphics[width=0.9\linewidth]{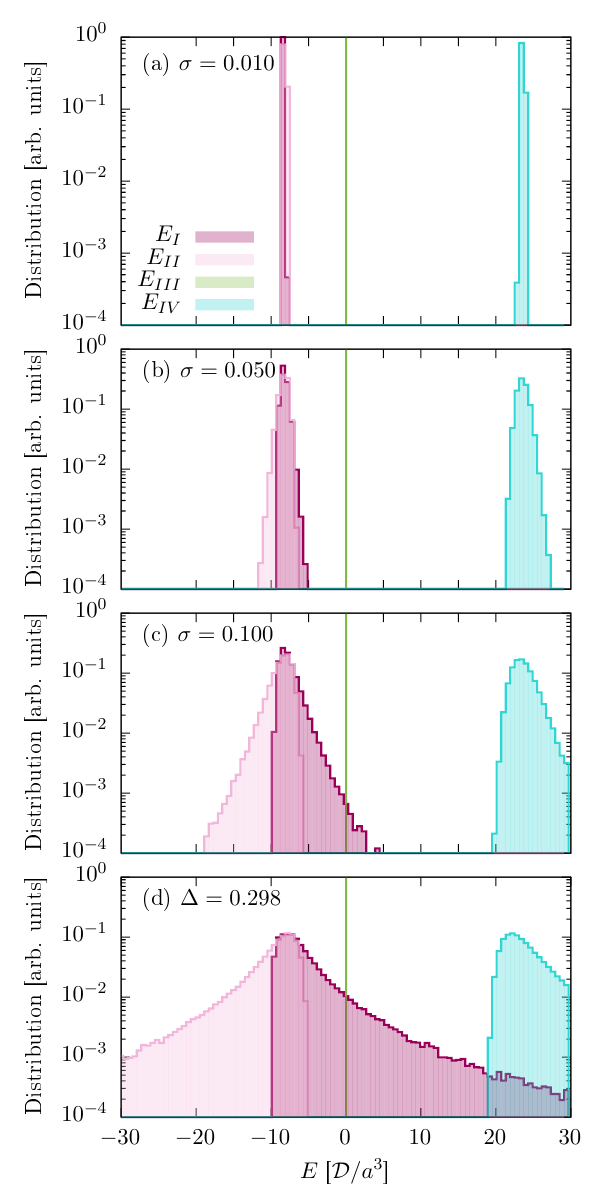}
\caption{Vertex energy spread for each topological group. Each graph corresponds to a different disorder strength. The distribution is in logarithmic scale so that the dispersion of the energies can be fully appreciated.}
\label{fig:spreadE}
\end{figure}

Note that each term of the energy of Type III vertices is zero (Eq.~\eqref{eq:energia3}).
Using Eqs.~\ref{eq:energia1}, \ref{eq:energia2} and \ref{eq:energia4} and sampling over the possible values of $d_\kappa$, we can calculate the dispersion of the vertex energies due to disorder.
In Fig.~\ref{fig:spreadE}, we show the results obtained. It can be observed how as disorder strength increases, the dispersion in the energies increases as well, which was expected. But, what is most interesting is to see how it increases.
As disorder increases, the energy distributions begin to overlap,
which would allow to obtain the ice state we mentioned in the previous section by adjusting the value of $\sigma$.
Then, while for Type I vertices the tail in the distribution extends to positive energy values, for Type II vertices this does it towards negative values. This causes Type II vertices to be more likely and increase their population as seen in Fig.~\ref{fig:verticespop}.
Another interesting result is that we can see why disorder does not allow defects to remain at low temperatures as could be expected.
The energy of Type III vertices is not affected by disorder since it is always zero and, in addition, the dispersion in the energy of Type IV vertices is always greater than that of other vertices and even their tail grows towards positive values.

On the other hand, to estimate the values obtained in the simulations for the population of Type I and II vertices at low temperatures (Fig.~\ref{fig:verticespop}), we can use the expressions found for the energies and perform the following algorithm.
Given a vertex with islands of lengths $\{d_1,d_2,d_3,d_4\} $, if the energy $E_\mathrm{I}(d_1, d_2, d_3, d_4)$ (Eq.~\eqref{eq:energia1}) is smaller than $E_\mathrm{II}(d_1, d_2, d_3, d_4)$ (Eq.~\eqref{eq:energia2}), then that vertex is of Type I and the number of Type I vertices, $ N_I $, increases by $1$.
If, on the other hand, $E_\mathrm{I}>E_\mathrm{II}$, then $N_{II}$ increases by $1$.
If we repeat this enough times, where the configuration $\{d_1, d_2, d_3, d_4\}$ is obtained according to the distribution and intensity of the disorder we want to analyze, we can estimate the value of $N_I$ and $N_{II}$ in the ground state for each value of $\sigma$.
Performing this algorithm for the parameters of Fig.~\ref{fig:spreadE}, we obtain the results shown in the Table~\ref{tab:vertexenergy}, and we found that the energetic analysis, regardless of its simplicity, manages to estimate the results from the MC numerical simulation, with a specially good match for $\sigma=0.05$.

\begin{table}
	\centering
	\begin{tabular}{@{}p{0.081\textwidth}*{4}{L{\dimexpr0.1\textwidth-2\tabcolsep\relax}}@{}}
		\toprule
		& \multicolumn{2}{c}{$N_{I}$ [\%]} &
		\multicolumn{2}{c}{$N_{II}$ [\%]} \\
		\cmidrule(r{4pt}){2-3} \cmidrule(l){4-5}
		Disorder strength
		& energetic analysis & numerical simulation & energetic analysis & numerical simulation \\
		\midrule
        $\sigma=0.000$ & 100 & 100 & 0 & 0\\
        $\sigma=0.010$ & 84.7(2) & 96.6(6) & 15.3(2) & 3.4(6)\\
        $\sigma=0.050$ & 51.0(2) & 50.3(5) & 49.0(2) & 49.7(5)\\
        $\sigma=0.100$ & 41.0(2) & 38.1(1) & 59.0(1) & 61.9(1)\\
        $\Delta=0.298$ & 30.3(1) & 25.3(4) & 69.7(2) & 74.8(4)\\
		\bottomrule
	\end{tabular}
	\caption{Estimated vertex population according to the energetic analysis described in Sec.~\ref{sec:vertexenergy}, compared with the lowest-temperature vertex population obtained from numerical simulations.}
    \label{tab:vertexenergy}
\end{table}

\begin{figure}
\centering
\includegraphics[width=0.95\linewidth]{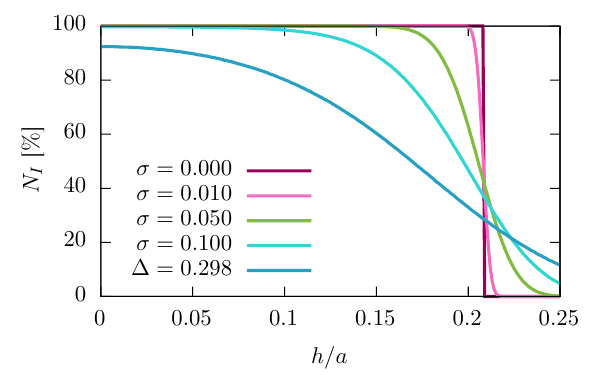}
\caption{Estimated population of Type I vertices for different disorder strengths as a function of the height parameter, obtained according to the algorithm described in Sec.~\ref{sec:vertexenergy} for the energetic analysis.}
\label{fig:popvsh}
\end{figure}

Using this simple algorithm, it also possible to show in detail the ferromagnetic-antiferromagnetic transition with the geometrical parameters.
In Fig.~\ref{fig:popvsh}, for $\sigma=0$, we show how the population for Type I vertices changes abruptly from $100$\% (antiferromagnetic state) to $0$\% (ferromagnetic state) while increasing the height parameter $h$.
We found that quenched disorder yields a rounding in this transition.
Samples with different values for $h$ have been manufactured experimentally and
the transition observed when measuring the vertex population for each sample is not abrupt as the non-disordered model predicts but it is in fact rounded~\cite{perrin2016extensive}.

\subsection{Slow dynamics}

We characterize the spin dynamics by calculating the spin-spin autocorrelation function
\begin{equation}
 C_{SS}(t) = \langle S_{i}(0) S_{i}(t) \rangle,
\end{equation}
where the initial orientation of each spin is compared with its orientation at time $t$ and brakets indicate an average over the $N$ sites of the lattice. In this way, we have that, at the beginning, $C_{SS}(0)=1$ and, as time passes and the system evolves, $C_{SS}(t)$ decreases, indicating the decorrelation and memory loss of the original state.

In the simplest case, the functional behavior of $C_{SS}(t)$ is given by an exponential $e^{-t/\tau}$ or stretched exponential $e^{-(t/\tau)^\beta}$~\cite{Takano1995,Zhou2017}, where $\tau$ is the relaxation time and $\beta$ is an exponent that takes values between $0$ and $1$.
The stretched exponential behavior is usually understood as the superposition of several exponential relaxations, each with a different decay~\cite{Phillips1996}.
However, there are some cases where $C_{SS}(t)$ can not be fitted with an exponential function; ours is one of them.
To estimate the time from which we can assume decorrelation, we define $\tau_0$ as the minimum time for which $C_{SS}(t)=0.00\pm0.01$ for all $t>\tau_0$,
where the error used in this definition is linked to fluctuations in our simulations.
According to this definition, $\tau_0$ will be systematically bigger than $\tau$, but it will be an useful estimator for our purposes.

\begin{figure}
\centering
\includegraphics[width=\linewidth]{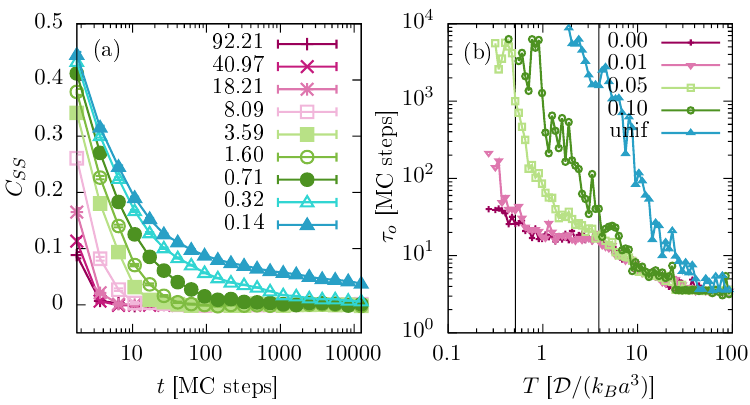}
\caption{(a) Spin-spin autocorrelation $C_{SS}(t) = \langle S_{i}(0) S_{i}(t) \rangle$ as a function of MC steps for different temperatures; simulation parameters: $N=576$, $r_{cut}=4$, $a=1$, $h=0.205$, $d=0.702$ and $\sigma=0.05$. (b) Relaxation time as a function of the temperature $T$ for different disorder strengths. Vertical lines indicate the crossover and phase transition temperatures for $\sigma=0$.}
\label{fig:autocorr}
\end{figure}

In Fig.~\ref{fig:autocorr}~(a), we show the spin-spin autocorrelation $C_{SS}(t)$ for $\sigma=0.05$. Different curves correspond to different temperatures.
It can be seen that, as we lower the temperature, the system slows down its dynamics.
Note that, to measure $C_{SS}$, the Parallel Tempering algorithm must be off (see Appendix~\ref{app:compdetails} for computational details).

Using this curves, we can calculate $\tau_0$ for each temperature; we show these results in Fig.\ref{fig:autocorr}~(b).
We can observe how, as disorder strength increases, low temperature dynamic freezes, losing ergodicity.
The energy map is full of local minima induced by disorder and these results are an indicator of how the system gets trapped in one of those.

\section{Conclusions}
\label{sec:conclusions}

We studied the effect of disorder on the geometry of an artificial spin ice with dipolar interactions.
First, we analyzed how disorder affects first and second nearest-neighbor bonds, and we selected appropriate geometrical parameters as well as the disorder intensities for our study.
Then, we studied the thermodynamic regimes for different disorder strengths.
Without disorder, the ground state of this system can be ferromagnetic or antiferromagnetic depending on the island lengths $d$.
Specifically, when $d\lesssim0.702$, $100\%$ of the vertices are of Type I (antiferromagnetic); but, if we increase $d$, the ground state will now be $100$\% Type II vertices (locally ferromagnetic). This ferromagnetic-antiferromagnetic transition with $d$, evidenced by a peak at the specific heat, disappears when disorder is added in the island lengths.
Instead, the ground state is formed by a mix of Type I and II vertices.
An important result is that this means that disorder strength can be chosen and tuned to thermalize the ice regime, where Type I and II vertices are equally likely.
Geometrical disorder shows interesting intermediate regimes such as the $50$--$50$\% ground state for $\sigma=0.05$, or, for strong disorder, a regime for which approximately $3/4$ of the vertices are Type II and $1/4$ are Type I.

To explain this behavior, we analyzed how disorder affects the spread of vertex energies.
To calculate this energy, we considered a single vertex and computed the interactions between each of its four islands.
By doing this, we found an asymmetry in this distribution that makes Type II vertices more likely.
The tail of this distribution moves towards negative values, while the tail of the distribution of Type I vertices moves towards positive values.
This also allows to understand why disorder does not permit defects at low temperatures in the needle model, contrary to expectations.
Also, with this analysis, we show in detail how the ferromagnetic-antiferromagnetic transition with $h$ is rounded by quenched disorder.
Finally, we found that disorder in the geometry causes a slowdown in the dynamics of the system which increases with disorder.

\appendix
\section{Computational details}
\label{app:compdetails}

We performed numerical simulations of the described system using the Monte Carlo (MC) method and the single-spin-flip~\cite{landau2005guide} and short-loop-move algorithms.
The latter is used to avoid the characteristic low-temperature freezing~\cite{Barkema1998}\cite[p. 143--148]{Moller2006t}.
At low temperatures, the vertices that do not meet the ice rule disappear (called ``defects'' in this context), leaving only Type I and II vertices. Flipping a single spin in these circumstances implies making a defect appear, which is unfavorable energetically.
As a consequence, the ocurrence of the single-spin-flip algorithm drops significantly, freezing the system.
The short-loop-move algorithm consists of finding a chain of spins and flipping them all together at once.
This process keeps the energy constant but allows access to a different configuration, thus recovering ergodicity.

We call MC step to $N$ iterations of the single-spin-flip algorithm, where $N$ is the size of the system, plus $N_{loop}$ iterations of the short-loop-move algorithm. The value of $N_{loop}$ is chosen to maximize the efficiency of the simulation.
The energy change is calculated according to Eqs.~\eqref{eq:hamiltoniano} and \eqref{eq:dip2}, and we use a cut-off distance for the dipolar sum.
According to previous works~\cite{Moller2006,Moller2006t,Moller2009}, it is known that the value for the temperature of the phase transition which appears in this system depends on this distance; if one were interested in calculating a more precise value for this transition temperature, it would be necessary to perform an Ewald sum.
Also, we assume periodic boundary conditions.

The use of quenched disorder also makes it necessary to use the Parallel Tempering (PT) method~\cite{Swendsen1986,Earl2005}.
This consists of running simultaneously $N_{PT}$ copies of the system at different temperatures. This copies, called replicas, are initialized randomly. At the end of each MC step, the different temperature configurations are switched according to Metropolis algorithm.
In this way, the high temperature configurations become accessible at low temperature and vice versa, accelerating considerably the dynamics and improving the ergodicity of the process.
The amount of replicas $N_{PT}$ must be chosen in such a way that the method has high acceptance, thus guaranteeing that all temperature configurations can be exchanged.
We chose the parameters to ensure an acceptance greater than $40\%$ at all temperatures.
We used this algorithm in all numerical simulations of this paper, except when spin-spin autocorrelation is measured.

We performed thermal and disorder average to calculate the values of the thermodynamic observables.
To ensure equilibrium, we measured the energy autocorrelation and we found that at least $10^6$ MC steps are necessary when $\sigma>0.05$, while $10^3$ MC steps is enough for weaker disorder.

The program was implemented in C++ using the Thrust library~\footnote{Thrust library: http://thrust.github.io/}, which allows parallelizing the calculation and running the same code in both GPUs and CPU's multicore, and it is available on demand.

\section*{Acknowledgments}
This work was partially supported by Consejo Nacional de Investigaciones Cient\'{\i}ficas y T\'ecnicas (CONICET), Argentina, PIP $\mathrm{N^o}\!$ 112-201301-00629.
This work used computational resources from CCAD -- Universidad Nacional de C\'ordoba~\footnote{CCAD -- UNC: http://ccad.unc.edu.ar/}, in particular the Mendieta Cluster, which is part of SNCAD -- MinCyT, Rep\'ublica Argentina.
The authors acknowledge D.~A. Martin, M.~L. Hoyuelos, J.~L. Iguain and P.~C. Guruciaga for fruitful discussions and suggestions.

\bibliography{geo-dis.bib}

\end{document}